\documentstyle[12pt]{article}
\newcommand{\be}{\begin{equation}}
\newcommand{\bea}{\begin{eqnarray}}
\newcommand{\eea}{\end{eqnarray}}
\newcommand{\ee}{\end{equation}}
\newcommand{\lb}{\label}
\newcommand{\f}{\frac}
\newcommand{\h}{{\cal H}}
\begin{document}

\begin{titlepage}
\begin{flushright}
Freiburg THEP-95/1\\
gr-qc/9501001
\end{flushright}
\vskip 1cm
\begin{center}
{\large\bf  QUANTUM GRAVITATIONAL EFFECTS IN} \\
 {\large\bf DE SITTER SPACE}\footnote{Invited Contribution to
 {\em New Frontiers in Gravitation}, edited by G. Sardanashvily
 and R. Santilli (Hadronic Press, 1995).}
\vskip 1cm
{\bf Claus Kiefer}
\vskip 0.4cm
 Fakult\"at f\"ur Physik, Universit\"at Freiburg,\\
  Hermann-Herder-Str. 3, D-79104 Freiburg, Germany.
\end{center}
\vskip 2cm
\begin{center}
{\bf Abstract}
\end{center}
\begin{quote}
We calculate the first quantum gravitational
 correction term to the trace anomaly in
De Sitter space from the Wheeler-DeWitt equation. This is obtained
through an expansion of the full wave functional for gravity and
a conformally coupled scalar field in powers of the Planck mass.
We also discuss a quantum gravity induced violation of unitarity
and comment on its possible relevance for inflation.
 \end{quote}

\end{titlepage}

A central role in the study of quantum field theory on a given classical
background spacetime is played by the semiclassical Einstein equations
\be R_{\mu\nu}-\f{1}{2}g_{\mu\nu}R =-8\pi G\langle T_{\mu\nu}
    \rangle, \lb{sc} \ee
in which the renormalised expectation value of the energy-momentum
tensor acts as a ``back reaction" on the metric of some classical
spacetime. The quantum matter state with respect to which this
expectation value is taken is assumed to obey, in the Schr\"odinger
picture, a functional Schr\"odinger equation, where the time evolution
is generated by the matter Hamiltonian. A prominent example is the case
of a conformally coupled scalar field in De~Sitter spacetime. If one
assumes this field to be in the Bunch-Davies vacuum state (which is
the unique De~Sitter invariant vacuum state \cite{FHJ}), one finds
for the expectation value of the energy-momentum tensor the result
(see e.g. \cite{BD})
\be \langle T_{\mu\nu}\rangle
 = \f{H_0^4\hbar}{960\pi^2}g_{\mu\nu}, \lb{em} \ee
where $H_0$ is the (constant) Hubble parameter of De~Sitter space.
Since the trace of this expression is non-vanishing, it leads to the
so-called {\em trace anomaly} because conformal invariance would lead
to a vanishing trace at the classical level.

Since one expects that the gravitational field is fundamentally described
by quantum theory, (\ref{sc}) can at best hold approximately.
There have been many discussions in the literature which have
investigated the range of validity of the semiclassical Einstein
equations. Ford \cite{Fo}, e.g., has compared the emission of
classical gravitational waves in the semiclassical theory with
graviton emission in linear quantum gravity and found that this can
be drastically different except if the mean deviation of
$T_{\mu\nu}$ is small. This is fulfilled, for example, if the
quantum state approximately evolves adiabatically. A number of papers
have strengthened this result by the attempt to derive (\ref{sc})
from the Wheeler-DeWitt equation, the central equation of
canonical quantum gravity, in a semiclassical approximation
and to study the emergence of the back reaction term with the help of
Wigner's function (see, e.g., \cite{PS}). A similar discussion
was made for quantum electrodynamics \cite{dec}.

If it is possible to derive the limit of quantum field theory
in a classical background from the Wheeler-DeWitt equation
\cite{BH}, one should also be able to go beyond this limit and
study quantum gravitational corrections. This has been achieved
through an expansion with respect to the gravitational constant
\cite{BH,KS}, although in a formal sense only, since regularisation
issues have not been adressed. More precisely, one obtains
correction terms to the functional Schr\"odinger equation for
matter fields on a given spacetime -- the first quantum gravity - induced
``post-Schr\"odinger approximation".

In the present paper these correction terms are explicitly
calculated for a conformally coupled scalar field in De~Sitter space.
Since the general discussion in \cite{KS} only dealt with
minimally coupled fields, it is necessary to extend the
approximation scheme to the case of conformally coupled fields.
This is nontrivial since the canonical formalism is very different
for non-minimally coupled fields \cite{Ki}.

The action for a massless scalar field $\phi$ which is conformally
coupled to gravity reads
\be S=\int d^4x\sqrt{-g}\left(\f{^{(4)}R -2\Lambda}{16\pi G}
   -\f{1}{2}g^{\mu\nu}\partial_{\mu}\phi\partial_{\nu}\phi
   -\f{1}{12}{^{(4)}}R\phi^2\right). \lb{ac} \ee
Canonical quantisation proceeds by first casting the theory into
Hamiltonian form. This is achieved by the usual 3+1 splitting
of spacetime into a foliation of spacelike hypersurfaces.
The Hamiltonian constraint found by this procedure is then
implemented in the usual way by applying its (naive) operator
version on physically allowed wave functionals, i.e., through the
Wheeler-DeWitt equation. It reads explicitly \cite{Ki}
\bea & & \h\Psi\equiv \left(\f{3\sqrt{h}}{16M}
  \f{\delta^2}{\delta(\sqrt{h})^2} -\f{1}{2M\sqrt{h}}
  \f{1}{1-\f{\phi^2}{24M}}\tilde{h}_{ac}\tilde{h}_{bd}
  \f{\delta^2}{\delta\tilde{h}_{ab}\delta\tilde{h}_{cd}}
  \right.\nonumber\\
  & & \ -\f{\phi}{8M}\f{\delta^2}{\delta\sqrt{h}\delta\phi}
  -\f{1}{2\sqrt{h}}\left(1-\f{\phi^2}{24M}\right)\f{\delta^2}
  {\delta\phi^2}-2\sqrt{h}MR \nonumber\\
  & & \ \ \left.+\f{\sqrt{h}\phi^2}{12}R +4\sqrt{h}M\Lambda
  +\f{\sqrt{h}}{2}h^{ab}\phi_{,a}\phi_{,b}\right)
  \Psi\left[\sqrt{h}({\bf x}),\tilde{h}_{ab}({\bf x}),
  \phi({\bf x})\right]=0,
  \lb{WDW} \eea
where $M=(32\pi G)^{-1}$, $h_{ab}=h^{1/3}\tilde{h}_{ab}$,
$h=\mbox{det}(h_{ab})$, and $R$ is the three-dimensional Ricci scalar.

We note that there exists a critical field value $\phi_c^2=
3/4\pi G$ for which the second term in (\ref{WDW}) diverges and
the signature of the kinetic term changes its sign. While this is
a crucial feature in full quantum gravity \cite{Ki}, it does
not affect the semiclassical expansion, as will be shown below.

The semiclassical expansion proceeds by writing
\be \Psi\equiv\exp(iS) \lb{psi} \ee
and expanding $S$ in powers of $M$:
\be S=MS_0+S_1+M^{-1}S_2+\ldots. \lb{S} \ee
This ansatz is inserted into (\ref{WDW}), and equal powers of $M$
are compared. The denominator in the second term is handled by
expanding
\be \left(1-\f{\phi^2}{24M}\right)^{-1} =1+
  \f{\phi^2}{24M}+\left(\f{\phi^2}{24M}\right)^2+\ldots, \lb{phi} \ee
and one recognises explicitly that the pole does not present
any problem in the semiclassical expansion.

The highest order in the semiclassical expansion is $M^2$,
and it leads to the condition that $S_0$ depend on the three-metric
only. The next order ($M$) yields the Hamilton-Jacobi equation
for gravity {\em alone},
\be -\f{3\sqrt{h}}{16}\left(\f{\delta S_0}{\delta\sqrt{h}}\right)^2
  +\f{1}{2\sqrt{h}}\tilde{h}_{ac}\tilde{h}_{bd}
  \f{\delta S_0}{\delta\tilde{h}_{ab}}\f{\delta S_0}
  {\delta\tilde{h}_{cd}}-2\sqrt{h}(R-2\Lambda)=0. \lb{HJ} \ee
Each solution of this equation describes a family of solutions
to the classical equations of motion (the vacuum Einstein equations),
i.e., a family of classical spacetimes. We shall choose a special
solution which corresponds to De~Sitter spacetime being foliated
into flat spatial slices, i.e., we consider regions of configuration
space where $R=0$. Looking thus for a solution of (\ref{HJ}) which depends
only on the three-dimensional volume, one finds
\be S_0=\pm 8\sqrt{\f{\Lambda}{3}}\int\sqrt{h}d^3x. \lb{S0} \ee
In the following we shall choose the solution with the minus sign.
It can be easily seen that (\ref{S0}) leads to an exponential
expansion for the scale factor $a$ ($\sqrt{h}\equiv a^3$), since
\be \f{\partial}{\partial t}\sqrt{h}\equiv
  \int d^3y\f{\delta\sqrt{h}({\bf x})}{\delta\tau({\bf y})}
     =\sqrt{3\Lambda h} \Rightarrow a(t)=e^{\sqrt{\f{\Lambda}{3}}t}
     \equiv e^{H_0t}. \lb{in} \ee
Note that the local ``WKB time" $\tau({\bf x})$ follows
from $S_0$ according to
\be \f{\delta}{\delta\tau({\bf x})}\equiv -\f{3\sqrt{h}}{8}
   \f{\delta S_0}{\delta\sqrt{h}}\f{\delta}{\delta\sqrt{h}}
   \equiv\sqrt{3h\Lambda}\f{\delta}{\delta\sqrt{h}}. \lb{tau} \ee
Note also that $a$ is chosen here to be dimensionless.

The next order ($M^0$) of our approximation scheme leads to an
equation involving $S_1$. Using the fact that $S_0$ does not depend
on $\tilde{h}_{ab}$, one obtains
\bea & & \f{3i\sqrt{h}}{16}\f{\delta^2 S_0}{\delta(\sqrt{h})^2}
  -\f{3\sqrt{h}}{8}\f{\delta S_0}{\delta\sqrt{h}}
  \f{\delta S_1}{\delta\sqrt{h}}+ \f{\phi}{8}
  \f{\delta S_0}{\delta\sqrt{h}}\f{\delta S_1}{\delta\phi}
  -\f{i}{2\sqrt{h}}\f{\delta^2 S_1}{\delta\phi^2} \nonumber\\
  & & \ +\f{1}{2\sqrt{h}}\left(\f{\delta S_1}{\delta\phi}
   \right)^2 +\f{\sqrt{h}\phi^2}{12}R +\f{\sqrt{h}}{2}
   h^{ab}\phi_{,a}\phi_{,b}=0. \lb{S1} \eea
We note that terms involving $\tilde{h}_{ab}$ begin to show up
only at order $M^{-2}$, which is one order beyond the orders
discussed in this paper. This demonstrates that the influence of
three-geometries which are not ``tangential" to curves in
configuration space is not seen in the present orders of approximation.
The expansion scheme thus proceeds as in the minimally coupled
case \cite{BH,KS}.

Writing $\psi\equiv D[\sqrt{h}]\exp(iS_1)$ and choosing the usual
prefactor equation for $D$ \cite{BH,KS}, (\ref{S1}) leads to
\be -i\f{3\sqrt{h}}{8}\f{\delta S_0}{\delta\sqrt{h}}
  \f{\delta\psi}{\delta\sqrt{h}}=\h_m\psi, \lb{TS} \ee
where $\h_m$ is the matter Hamiltonian density which reads explicitly
\be \h_m=-\f{1}{2\sqrt{h}}\f{\delta^2}{\delta\phi^2}
  -\f{i\phi}{8}\f{\delta S_0}{\delta\sqrt{h}}\f{\delta}{\delta\phi}
  +\f{\sqrt{h}}{2}\left(\f{\phi^2}{6}R+h^{ab}\phi_{,a}\phi_{,b}
  \right). \lb{Hm} \ee
The left-hand side of Eq. (\ref{TS}) is often written as
$i\delta\psi/\delta\tau$, see (\ref{tau}), but one must keep
in mind that this notion is misleading, since the presence of such
a time function would contradict the commutation relations between
the Hamiltonian densities at different space points \cite{GK}.
Anyway, the important equation for the following discussion
is the {\em integrated} version of (\ref{TS}) along a particular
choice of the slicing. This functional Schr\"odinger equation reads
\be i\f{\partial\psi}{\partial t}=\int d^3x
  \left(-\f{1}{2a^3}\f{\delta^2}{\delta\phi^2}+i\phi H_0
  \f{\delta}{\delta\phi} +\f{a}{2}(\nabla\phi)^2\right)\psi
  \equiv H_m\psi.
  \lb{FS} \ee
Since De Sitter space is homogeneous, it is convenient to introduce
the Fourier transform of $\phi(x)$,
\be \phi({\bf x})=\int\f{d^3k}{(2\pi)^3}\chi({\bf k})e^{i{\bf kx}}
  \equiv\int d^3\tilde{k}\chi_ke^{i{\bf kx}}. \lb{FT} \ee
Note that
\be \f{\delta}{\delta\phi({\bf x})}=\int d^3\tilde{k}
  e^{i{\bf kx}}\f{\delta}{\delta\chi_k} \Rightarrow
  \f{\delta\chi_k}{\delta\chi_{k'}}=(2\pi)^3\delta({\bf k}+
  {\bf k'}). \lb{der} \ee
Since $\phi$ occurs only quadratically in (\ref{FS}), it is suggesting
to make a Gaussian ansatz for the wave functional $\psi$,
\be \psi[\chi_k,t)= N(t)\exp\left(-\f{1}{2}\int d^3\tilde{k}
   \Omega({\bf k},t)\chi_k\chi_{-k}\right). \lb{gau} \ee
Such a state describes a general vacuum state (independent of any
Fock space) for the scalar field in the gravitational background.
Inserting (\ref{gau}) into (\ref{FS}), one immediately obtains
two equations for $N$ and $\Omega$,
\bea i\f{\dot{N}}{N} &=& \f{V}{2a^3}\int d^3\tilde{k}\Omega
     \equiv\f{1}{2a^3}\mbox{Tr}\Omega, \lb{Om1} \\
     i\dot{\Omega} &=& \f{\Omega^2}{a^3}-ak^2 +2iH_0\Omega,
     \lb{Om2} \eea
 where $k\equiv\vert{\bf k}\vert$, and $V$ is the space volume which
 is introduced for regularisation. Given an initial state $\psi_0$, these
 equations uniquely determine the state $\psi$ for all times
 (we assume the normalisation $\langle\psi\vert\psi\rangle=1$).
 Eq. (\ref{Om2}) can easily be solved by introducing a quantity
 $y$ according to
 \be \Omega=-ia^3\f{\dot{y}}{y} \lb{defy} \ee
 which then leads to a linear equation for $y$,
 \be \ddot{y}+H_0\dot{y}+\f{k^2}{a^2}y=0. \lb{yt} \ee
 Introducing the conformal time coordinate $\eta$ according to
 \be dt=a(\eta)d\eta \Rightarrow a(\eta)=-\f{1}{H_0\eta},\;
     \eta\in(0,-\infty), \lb{con} \ee
 and denoting derivatives with respect to $\eta$ by primes,
 this reduces to the simple form
 \be y''+k^2y=0. \lb{yn} \ee
 We choose the solution
 \be y(\eta)=\f{1}{\sqrt{2k}}e^{ik\eta} \lb{y} \ee
 which selects the positive frequency solution $\Omega
 =\mbox{Re}\Omega=a^2k$. Eq. (\ref{y}) corresponds to the conformal
 vacuum state which is known to agree with the adiabatic vacuum state
 in the massless limit \cite{BD}.

One important point has to be emphasised. Our matter Hamiltonian
(\ref{Hm}) {\em differs} from the corresponding Hamiltonian
density which is obtained by directly inserting a solution
of the classical Einstein equations into (\ref{ac}) \cite{GLH}.
The reason for this discrepancy is the non-minimal coupling
of $\phi$ to $R$, which couples kinetic gravitational terms
to the matter field. It thus makes a difference whether gravity is
described by quantum theory at a fundamental level and a
semiclassical approximation is performed, or whether gravity is
treated classically ab initio.

Proceeding with the expansion scheme to the next order
($M^{-1}$), one finds the same correction terms to the
Schr\"odinger equation than in the minimally coupled case
\cite{BH,KS}. The
``corrected Schr\"odinger equation" thus reads
(re-inserting $\hbar$)
\be i\hbar\f{\partial\psi}{\partial t}=H_m\psi
    -\f{2\pi G}{\Lambda a^3}\int d^3x\h_m^2\psi -i\hbar
    \f{2\pi G}{V\Lambda a^3}\left(\f{\partial}{\partial t}
    H_m\right)\psi. \lb{corr} \ee
In the following we shall use these correction terms to evaluate
the quantum gravity-induced correction to the expectation value
(\ref{em}). To this purpose we need the expectation value of
the Hamiltonian density $\h_m$ with respect to the Gaussian
state (\ref{gau}). With the explicit form given in (\ref{Hm}), one
finds\footnote{We also take into account an imaginary part for
$\Omega$, $\Omega_I$, since this is necessary for the dimensional
regularisation of the expectation values (in dimensions away from
3 space dimensions, $\Omega$ is {\em complex})}
\be \langle\h_m\rangle= \f{1}{4a^3}\int d\tilde{k}
    \f{\vert\Omega\vert^2}{\Omega_R} +\f{H_0}{2}
    \int d\tilde{k}\f{\Omega_I}{\Omega_R}
    +\f{a}{4}\int d\tilde{k}\f{k^2}{\Omega_R},
    \lb{exh} \ee
 where $d\tilde{k}\equiv dk/(2\pi)^3$. Note that
 \be \langle\phi^2\rangle =\int\f{d\tilde{k}}{2\Omega_R}
   =2\pi H_0^2\eta^2\int_0^{\infty}d\tilde{k}k
   =\f{H_0^2}{4\pi^2}\int_0^{\infty}udu, \lb{exph} \ee
which is formally independent of conformal time $\eta$ (this
is a property of the special Gaussian state chosen). This, as well as
\[ \langle\h_m\rangle =\f{H_0^4}{4\pi^4}\int_0^{\infty}
   u^3du, \]
is of course ultraviolet divergent. One thus needs a regularisation
prescription to arrive at a physically sensible, finite, result.
For ultraviolet regularisation the high frequency modes in
(\ref{gau}) are important. For such modes the Hamiltonian density in
\cite{GLH} coincides with our Hamiltonian
density. One can thus use the
result for $\langle T_{\mu\nu}\rangle$ already found in
\cite{GLH}. However, one must keep in mind that for non-minimally
coupled fields $T_{00}$ does {\em not} coincide with the
Hamiltonian density. In fact, one finds that
\be a^2\langle T_{00}\rangle -\langle\h_m\rangle
   =\f{a^3H_0^2}{4}\int\f{d\tilde{k}}{\Omega_R}
   -H_0\int d\tilde{k}\f{\Omega_I}{\Omega_R}. \lb{diff} \ee
Using (\ref{Om2}) as well as the results of \cite{GLH}, one has
\[ \langle\phi^2\rangle= -\f{1}{a^3}\f{\langle\h_m\rangle}
   {H_0^2} \]
   and
 \[ \langle\h_m\rangle =-\f{2}{3}a^3\langle T_{00}\rangle. \]
 Using (\ref{em}), one eventually finds
 \be \langle\h_m\rangle =\f{a^3\hbar H_0^4}{1440\pi^2}.
     \lb{exp} \ee
Considering now the quantum-gravity induced correction terms
to the functional Schr\"odinger equation, given in (\ref{corr}),
one obtains from the first term a shift $\delta\epsilon$
to the expectation value (\ref{exp}). The second term describes
a quantum gravity induced unitarity violation and will be discussed
below. The shift $\delta\epsilon$ is then given by
\be \delta\epsilon =-\f{2\pi G}{3a^3H_0^2}\langle\h_m^2\rangle.
   \lb{sh} \ee
This expression is in general utterly divergent and has to be
regularised. In the present case, however, the state evolves adiabatically,
i.e., the mean deviation of $\h_m$ can be assumed to be small:
$\langle\h_m^2\rangle\approx\langle\h_m\rangle^2$. As was mentioned above,
this is the condition under which the semiclassical Einstein equations
(\ref{sc}) are valid. This fact, together with (\ref{exp}),
immediately leads to the result
\be \delta\epsilon\approx -\f{2\pi G}{3a^3H_0^2}\langle
   \h_m\rangle^2 =-\f{2G\hbar^2H_0^6a^3}{3(1440)^2\pi^3}.
   \lb{shres} \ee
It is clear that, due to the presence of $G\hbar^2$, this shift
is in general very small. We emphasise that (\ref{shres}) is a
definite prediction from the Wheeler-DeWitt equation.

We now turn to the discussion of the second
correction term in (\ref{corr}).
It contributes an imaginary part $\epsilon_I$ to the energy density,
which might be interpreted as an instability of the vacuum matter
state due to the emission of gravitons. This unitarity violation
can be understood from the fact that the Wheeler-DeWitt equation
does not obey a conservation law for a Schr\"odinger current, but
for a Klein-Gordon current \cite{BH}.
The imaginary contribution to the energy reads
\be \epsilon_I= -\f{2\pi G\hbar}{3a^3VH_0^2}
    \left\langle\f{\partial\h_m}{\partial t}\right\rangle.
    \lb{EI1} \ee
This can be evaluated by using the explicit expression
(\ref{Hm}) and calculating expectation values with respect to
the state (\ref{gau}) in the same manner as above.
One finds that
\[    \left\langle\f{\partial\h_m}{\partial t}\right\rangle
      =3H_0\langle\h_m\rangle, \]
leading to
\be \epsilon_I= -\f{G\hbar^2H_0^3}{720\pi V}. \lb{EI2} \ee
The time scale $t^*$ of this quantum gravitational instability
(the sign of $\epsilon_I$ signals that it is a {\em decay})
is thus given by
\be t^*= \hbar\vert 2\epsilon_IV\vert^{-1}
       =\f{360\pi}{G\hbar H_0^3}\sim \left(\f{H_0^{-1}}{t_P}
       \right)^3t_P, \lb{ts} \ee
 where $t_P$ is the Planck time. This time scale is thus huge,
 except if $H_0^{-1}$ is of the order of the Planck length.
 Nevertheless, this correction term might have some relevance
 for inflationary cosmology. The reason is that in the early universe,
 during inflation, $H_0$ is supposed to be large, so that the
 time scale $t^*$ may be small enough to become relevant.
 Specifically, if
 \[ 10^{-34}t_P^{-1}\leq H_0\leq 10^{-6}t_P^{-1}, \]
 then
 \[ 10^{18}t_P\leq t^*\leq 10^{102}t_P. \]
 Thus, $t^*$ may be small enough to enable a ``quantum gravity induced"
 decay of the matter state, which, in turn, may produce a mechanism
 for a natural exit from the inflationary stage of the universe.
 We note that it has also been concluded from a two-loop calculation
 of pure gravity that infrared effects may be responsible for such
 a natural exit \cite{Wo}.

We finally mention that the back reaction of the matter state onto
the gravitational background leads to a change in the definition
of semiclassical time and thus to an additional contribution
to the corrected Schr\"odinger equation (\ref{corr}) \cite{BH}.
The net effect of this additional term is a {\em change of sign}
in the energy shift (\ref{shres}) \cite{BH}, and there is no change
in the unitarity violating term (\ref{EI2}).

In conclusion, we have calculated definite predictions of
quantum gravity from the Wheeler-DeWitt equation. It should be
straightforward to apply this approach to other situations where
the matter state evolves adiabatically with respect to the
background \cite{BD}. In the general situation, however,
the corrections terms in (\ref{corr}) must be regularised carefully.
We also note that the unitarity violating contribution becomes
relevant in the final stages of black hole evaporation
\cite{KMS}.

\end{document}